%% file: jades_bd.tex
\documentclass{aa}

\usepackage{newtxtext,newtxmath}

\usepackage[T1]{fontenc}

\DeclareRobustCommand{\VAN}[3]{#2}
\let\VANthebibliography\thebibliography
\def\thebibliography{\DeclareRobustCommand{\VAN}[3]{##3}\VANthebibliography}

\usepackage{graphicx}	
\usepackage{amsmath}	
\usepackage{textgreek}
\usepackage{xspace}
\input{config/macros}

\newcommand{\um}{\text{\textmu m}\xspace}

\newcommand{\target}{JADES-GS+53.16904-27.77884\xspace} 
\newcommand{\ppxf}{\textsc{ppxf}\xspace} 
\newcommand{\beagle}{\textsc{beagle}}

\begin{document}

\title{JADES: Balmer Decrement Measurements at redshifts $4<z<7$}


    \author{Lester Sandles\inst{\ref{Kavli},\ref{Cavendish}}\thanks{E-mail: ls861@cam.ac.uk} 
    \and Francesco D'Eugenio\inst{\ref{Kavli},\ref{Cavendish}}
    \and Roberto Maiolino\inst{\ref{Kavli},\ref{Cavendish},\ref{RM3}}
    \and Tobias J.\ Looser\inst{\ref{Kavli},\ref{Cavendish}}
    \and Santiago Arribas\inst{\ref{SA}}
    \and William M.\ Baker\inst{\ref{Kavli},\ref{Cavendish}}
    \and Nina Bonaventura\inst{\ref{NB},\ref{NB2},\ref{Arizona}}
    \and Andrew J.\ Bunker\inst{\ref{Oxford}}
    \and Alex J.\ Cameron\inst{\ref{Oxford}}
    \and Stefano Carniani\inst{\ref{SCa}}
    \and Stephane Charlot\inst{\ref{SCh}}
    \and Jacopo Chevallard\inst{\ref{Oxford}}
    \and Mirko Curti\inst{\ref{MK},\ref{Kavli},\ref{Cavendish}}
    \and Emma Curtis-Lake\inst{\ref{ECL}}
    \and Anna de Graaff\inst{\ref{Max}}
    \and Daniel J.\ Eisenstein\inst{\ref{Harvard}}
    \and Kevin Hainline\inst{\ref{Arizona}}
    \and Zhiyuan Ji\inst{\ref{Arizona}}
    \and Benjamin D.\ Johnson\inst{\ref{Harvard}}
    \and Gareth C.\ Jones\inst{\ref{Oxford}}
    \and Nimisha Kumari\inst{\ref{NK}}
    \and Erica Nelson\inst{\ref{Colorado}}
    \and Michele Perna\inst{\ref{SA}}
    \and Tim Rawle\inst{\ref{TR}}
    \and Hans-Walter Rix\inst{\ref{Max}}
    \and Brant Robertson\inst{\ref{BR}}
    \and Bruno Rodr\'iguez Del Pino\inst{\ref{SA}}
    \and Jan Scholtz\inst{\ref{Kavli},\ref{Cavendish}}
    \and Irene Shivaei\inst{\ref{Arizona}}
    \and Renske Smit\inst{\ref{RS}}
    \and Fengwu Sun\inst{\ref{Arizona}}
    \and Sandro Tacchella\inst{\ref{Kavli},\ref{Cavendish}}
    \and Hannah \"Ubler\inst{\ref{Kavli},\ref{Cavendish}}
    \and Christina C.\ Williams\inst{\ref{CWilliams}}
    \and Chris Willott\inst{\ref{CWillott}}
    \and Joris Witstok\inst{\ref{Kavli},\ref{Cavendish}}
}

\institute{Kavli Institute for Cosmology, University of Cambridge, Madingley Road, Cambridge CB3 0HA, UK\label{Kavli}
\and Cavendish Laboratory, University of Cambridge, 19 JJ Thomson Avenue, Cambridge CB3 0HE, UK\label{Cavendish}
\and Department of Physics and Astronomy, University College London, Gower Street, London WC1E 6BT, UK\label{RM3}
\and Centro de Astrobiolog\'ia (CAB), CSIC–INTA, Cra. de Ajalvir Km.~4, 28850- Torrej\'on de Ardoz, Madrid, Spain\label{SA}
\and Cosmic Dawn Center (DAWN), Copenhagen, Denmark\label{NB}
\and Niels Bohr Institute, University of Copenhagen, Jagtvej 128, DK-2200, Copenhagen, Denmark\label{NB2}
\and Steward Observatory, University of Arizona, 933 N. Cherry Avenue, Tucson AZ 85721 USA\label{Arizona}
\and Department of Physics, University of Oxford, Denys Wilkinson Building, Keble Road, Oxford OX1 3RH, UK\label{Oxford}
\and Scuola Normale Superiore, Piazza dei Cavalieri 7, I-56126 Pisa, Italy\label{SCa}
\and Sorbonne Universit\'e, CNRS, UMR 7095, Institut d'Astrophysique de Paris, 98 bis bd Arago, 75014 Paris, France\label{SCh}
\and European Southern Observatory, Karl-Schwarzschild-Strasse 2, 85748 Garching, Germany\label{MK}
\and Centre for Astrophysics Research, Department of Physics, Astronomy and Mathematics, University of Hertfordshire, Hatfield AL10 9AB, UK\label{ECL}
\and Max-Planck-Institut f\"ur Astronomie, K\"onigstuhl 17, D-69117, Heidelberg, Germany\label{Max}
\and Center for Astrophysics $|$ Harvard \& Smithsonian, 60 Garden St., Cambridge MA 02138 USA\label{Harvard}
\and AURA for European Space Agency, Space Telescope Science Institute, 3700 San Martin Drive. Baltimore, MD, 21210\label{NK}
\and Department for Astrophysical and Planetary Science, University of Colorado, Boulder, CO 80309 USA\label{Colorado}
\and European Space Agency (ESA), European Space Astronomy Centre (ESAC), Camino Bajo del Castillo s/n, 28692 Villafranca del Castillo, Madrid, Spain\label{TR}
\and Department of Astronomy and Astrophysics University of California, Santa Cruz, 1156 High Street, Santa Cruz CA 96054 USA\label{BR}
\and Astrophysics Research Institute, Liverpool John Moores University, 146 Brownlow Hill, Liverpool L3 5RF, UK\label{RS}
\and NSF’s National Optical-Infrared Astronomy Research Laboratory, 950 North Cherry Avenue, Tucson, AZ 85719 USA\label{CWilliams}
\and NRC Herzberg, 5071 West Saanich Rd, Victoria, BC V9E 2E7, Canada\label{CWillott}
}
 
\authorrunning{L.\ Sandles et al.}
\date{}

\abstract{
We present Balmer decrement \Halpha / \Hbeta measurements for a sample of 51 galaxies at redshifts $z=4\text{--}7$ observed with the \JWST/NIRSpec MSA, as part of the JADES survey.
Leveraging 28-hour long exposures and the efficiency of the prism/clear configuration (but also using information from the medium-resolution gratings),
we are able to probe directly the low-mass end of the galaxy population, reaching
stellar masses \Mstar as low as $10^7$~\MSun. We find that the correlation between Balmer
decrement and \Mstar is already established at these high redshifts, indicating a rapid build up of dust in moderately massive galaxies at such early epochs. The lowest-mass
galaxies in our sample ($\Mstar = 1\text{--}3\times10^7$~\MSun) display a remarkably low Balmer decrement of $2.88\pm0.08$, consistent with case B, suggesting very little dust content. However, we warn that such a low observed Balmer decrement may also partly be a consequence of an intrinsically lower \Halpha / \Hbeta, resulting from the extreme conditions of the ionized gas in these primeval and unevolved systems.
We further compare the Balmer decrement to continuum-derived star-formation rates 
(SFR), finding tentative evidence of a correlation, which likely traces the underlying connection between SFR and mass of cold gas. However, we note that larger samples are required to distinguish between direct and primary correlations from indirect and secondary dependencies at such high redshifts.
}


\maketitle



\section{Introduction}

The Balmer recombination lines are an efficient probe of star-forming regions. They are the brightest non-resonant series
(enabling efficient escape, unlike the intrinsically brighter Lyman series) and they are less affected by dust attenuation compared to probes at shorter wavelengths.
The Balmer series is therefore the brightest measure of the flux of hydrogen-ionising continuum emitted by massive, short-lived ($<10$~Myr) stars.
These lines therefore provide the most direct measure of the current rate of star formation, on timescales of 3--10~Myr \citep{flores_velazquez_time-scales_2021, tacchella_h_2022}.

Because of the wavelength-dependent attenuation from dust, the observed flux ratio
between the Balmer lines (Balmer decrement) provides a direct measure of the optical depth towards star-forming regions. This measurement is essential to complement
alternative probes such as the attenuation of the optical and UV stellar continuum
\citep[e.g.,][]{galliano_interstellar_2018, salim_dust_2020}, which probe different optical depths \citep[e.g.,][]{1994ApJ...429..582C}.

While this measurement requires knowledge of the intrinsic
ratio, the latter value is known from atomic physics and is relatively insensitive
to a broad range of different physical conditions commonly found in star-forming regions \citep{Osterbrock_2006, smith_physics_2022}.
In particular, variations of the gas temperature between 5,000--20,000~K and of the density between 100--10,000~\percmq can change the Balmer decrement by less than 10~per cent.

Leveraging these advantages, the ratio of the two brightest lines -- \Halpha/\Hbeta -- has been used to estimate the dust content of star-forming regions \citep{reddy_mosdef_2015,nelson_spatially_2016,matharu_first_2023},
the amount of gas present \citep[e.g.,][]{piotrowska_towards_2020}, and their
relation with other galaxy properties \citep[e.g.,][]{groves_balmer_2012}. 
\citet{maheson_unravelling_2023} show that the Balmer decrement correlates with stellar mass \Mstar, with secondary dependencies on gas metallicity and velocity dispersion.
One should however also take into account that the geometry of the dust distribution also plays a role in the observed Balmer decrement. Specifically, very dusty and optically thick systems can potentially result in a low Balmer decrement, as the Balmer lines may trace only the outer, optically thin regions of the galaxy.

However, as we strive to observe galaxies further away, at earlier cosmic times, the Balmer series is redshifted towards near-infrared (NIR) wavelengths, where atmospheric absorption, as well as thermal and telluric emission from the atmosphere, make ground-based observations difficult. Despite that, in recent years,
ground-based NIR surveys have revealed how the Balmer decrement evolves up to Cosmic Noon, at redshifts $z=2\text{--}3$ \citep{kriek_mosfire_2015, reddy_mosdef_2015, curti_klever_2020, lorenz_updated_2023}.
Yet, at $z>3$, \Halpha is redshifted beyond 2.6~\um, where
ground-based observations of faint distant galaxies are prohibitively expensive.
This meant that, at these high redshifts, dust could only be probed using sub-mm observations \citep{watson_dusty_2015, 2017ApJ...837L..21L, 2022MNRAS.515.1751W}, which is feasible only for relatively high stellar masses.

The start of \JWST science operations in July 2022 opened for the first time a view into \Halpha at redshifts $z>3$ \citep{gardner_james_2023}. Early results using data from the Cosmic Evolution Early Release Science Survey
\citep[CEERS; ][]{finkelstein_ceers_2023} have shown that the correlation between Balmer decrement and stellar mass is already established at $z=4\text{--}6$ \citep{shapley_jwstnirspec_2023}. In particular, low-mass galaxies are confirmed to have low Balmer decrement -- suggesting low dust content.
At the same time, studies of collisionally excited lines have shown that the star-forming regions of early galaxies were hotter than in the local Universe, with electron temperatures of $\approx$20,000~K \citep[e.g.,][]{schaerer_first_2022, curti_chemical_2023, curti_jades_2023} and possibly denser, with electron densities $\nelec \approx 500$~\percmq \citep{reddy_jwstnirspec_2023}.
These differences mean that the intrinsic Balmer decrement should also be slightly different, requiring a correction to our methods to estimate dust attenuation.

In this work, we leverage the unprecedented depth of the \JWST Advanced Deep Extragalactic Survey \citep[JADES;][]{eisenstein_overview_2023} to probe dust attenuation in the star-forming regions of galaxies at redshifts $z=4\text{--}7$.
In \S~\ref{s.data} we introduce the data and sample and describe the measurements we use. Our results are presented in \S~\ref{s.results} and we conclude the article with a
discussion of the implications (\S~\ref{s.discussion}) and a summary
(\S~\ref{s.conclusions}).

Throughout this work, we assume a flat $\Lambda$CDM cosmology with $H_0 = 70$~km s$^{-1}$ Mpc$^{-1}$, $\Omega_{\textrm{M}} = 0.3$ and $\Omega_\Lambda = 0.7$. For the
initial mass function we assume the functional shape of \citet{chabrier_galactic_2003}.

\section{Data / Sample Selection}\label{s.data}

\begin{figure*}
    \centering
    \includegraphics[width=\textwidth]{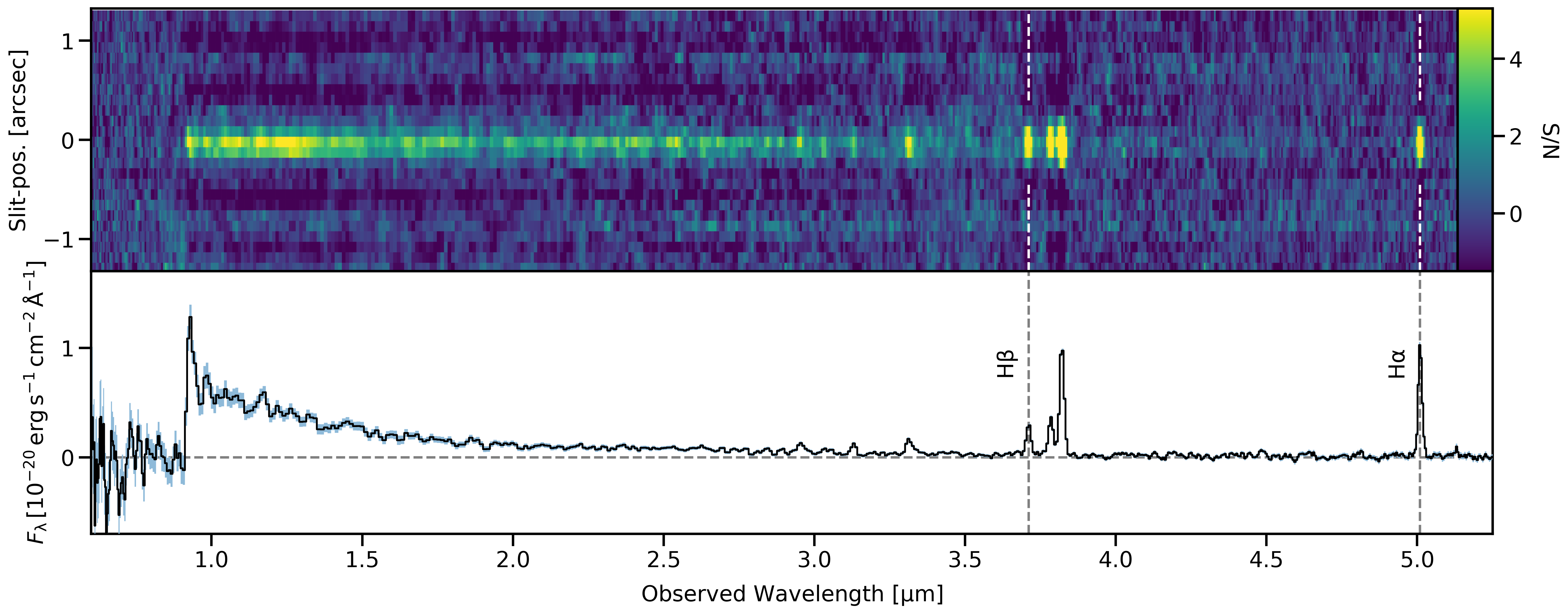}
    \caption{NIRSpec low-resolution spectrum of \target with a redshift of $z=6.6$. The top panel shows the 2D spectrum colour coded by the signal-to-noise ratio; S/N. The bottom panel shows the extracted 1D flux density versus observed wavelength.}
    \label{fig:spec.png}
\end{figure*}

\subsection{Data}

We use data from the JADES survey \citep{bunker_jades_2023, eisenstein_overview_2023}, obtained as part of
programme ID 1210 (PI: N. Lützgendorf). The targets were selected from the GOODS-South field \citep{giavalisco_great_2004}, as described in \citet{bunker_jades_2023}. \JWST/NIRSpec was configured to use the micro-shutter assembly \citep[MSA;][]{jakobsen_near-infrared_2022}, with the prism/clear disperser/filter combination to obtain low-resolution spectroscopy (nominal $R=30\text{--}100$) for 253 targets between $0.6 < \lambda < 5.3$~\um. An example spectrum (\target) is shown in Fig.~\ref{fig:spec.png}.

Each target was assigned to an array of $3\times 1$ micro-shutters, forming an effective slit of 0.2~arcsec width and $\approx$1.3~arcsec length. We used three dither positions and three nods, for a total integration time of up to 28~hours. The data reduction was performed using the pipeline developed by the ESA NIRSpec Science Operations Team and the NIRSpec GTO Team.
We use the publicly available data products v3.0, which include a wavelength dependent slit-loss correction, assuming all sources are unresolved.
We refer to Carniani et~al. (in~prep) and to \citep{cameron_jades_2023, curti_jades_2023, curtis-lake_spectroscopic_2023} for more information on the data reduction process.

In addition to prism spectroscopy, JADES also 
provides medium-resolution (nominal $R=1000$) spectroscopy covering the full 0.6--5.3~\um range,
combining the three disperser/filter configurations: g140m/f070lp, g235m/f170lp and g395m/f290lp. This dataset is less deep 
than the prism observations, owing to shorter exposure times 
($\approx$7~hours). Moreover, medium-resolution observations do not cover the entire sample.

\subsection{Target selection and redshift determination}

The target selection was based on the photometric Lyman-break technique, as described in \citet{bunker_jades_2023}. With the sensitivity of our observations, we probe stellar masses down to $10^6~\MSun$ (but in this work we are limited to $10^7~\MSun$ due to additional requirements in the emission-line signal-to-noise ratio; S/N).

Galaxy redshifts were obtained with an iterative procedure. We use \textsc{bagpipes} \citep{carnall_inferring_2018} to infer the spectroscopic redshift from the prism observations. This redshift is subsequently refined by visual inspection of the medium- and high-resolution spectra (if available) and by fitting the emission lines with a Gaussian. In this work, we always require detection of at least two emission lines (\Hbeta and \Halpha), therefore the redshifts are always determined well, within the accuracy of the spectroscopic data \citetext{\citealp{bunker_jades_2023}, Jakobsen et~al., in~prep}.

At redshifts $z<4$ the spectral resolution of the prism is unable to resolve the 0.1~\um gap between \Hbeta and \OIIIL[4959]. Noting that \Halpha is redshifted beyond 5.3~\um at redshift $z=7$, we apply a redshift cut of $4<z<7$, reducing our sample size to 58 (out of 253) objects.

\subsection{Emission-line measurements}
\label{ss:Emission-line measurements}

Emission-line fluxes are measured using simultaneous fitting of the stellar continuum and nebular emission with the software \ppxf \citep{cappellari_improving_2017, cappellari_full_2022}. 

The stellar continuum is modelled as a non-negative linear superposition of single stellar population templates (SSP). The SSP spectra use C3K synthetic model atmospheres \citep{conroy_resolving_2019} and MIST isochrones \citep{choi_mesa_2016, dotter_mesa_2016}, which we convolve with a wavelength-dependent Gaussian kernel corresponding to 0.5$\times$ the nominal spectral resolution of the NIRSpec prism. This is to provide a first-order correction to the increased spectral resolution for sources that are smaller than the micro-shutter width. Any flux blueward of \Lyalpha is set to zero, because we do not model foreground absorption due to neutral-gas. In addition to the SSP spectra, the continuum also includes a 5\textsuperscript{th}-order multiplicative Legendre polynomial, to model the effect of dust and the possible damping wing at rest-frame 0.12~\um.
In this work we focus on rest-frame 0.4--0.7~\um, where the effect of these modelling choices is negligible. 

Emission lines are modelled as Gaussians, with a reduced set of parameters to maximise S/N; in particular, for this study, we tie together the velocity and velocity dispersion of the \Hbeta--\OIIIL[4959,5007] complex and of the \Halpha--\NIIL[6548,6584] complex (hereafter, we always refer to the \OIIIL[4959,5007] and \NIIL[6548,6584] doublets as \OIII and \NII).
When fixed by atomic physics, the ratio of emission-line doublets is constrained, which in this paper is relevant for both the \OIII and \NII doublets. A critical aspect of these emission-line measurements is that -- by simultaneously modelling the stellar continuum -- our algorithm includes a correction for stellar-atmospheric absorption.

Using low-resolution instead of medium-resolution spectroscopy presents advantages and disadvantages.
In medium resolution, we always resolve the \Hbeta--\OIII and the \Halpha--\NII complexes, giving more accurate measurements. However, medium-resolution observations are less deep and cover a smaller sample than the low-resolution observations. This means using low-resolution increases the precision
at the expense of accuracy. 

Within this context, we note that the \Halpha and \NII lines are blended in the prism. Therefore, in order to measure \Halpha, we first measure the total $\Halpha+\NII$ flux from the prism.
For five objects in our sample, we detect \NII in the corresponding medium-resolution spectra (S/N$>3$). For these objects we multiply the total low-resolution $\Halpha+\NII$ flux by the medium-resolution corrective factor: $\Halpha / (\Halpha+\NII)$. For the remaining objects, we use a constant correction of $\Halpha / (\Halpha+\NII) = 0.96$ determined from the stack of all $z>4$ medium-resolution spectra. 

From our sample of 58 galaxies ($4<z<7$), we remove seven objects with \Hbeta S/N $<3$: three of which also had an \Halpha S/N $<3$; one of which has a `blocky', asymmetric \Halpha profile (probably resulting from a noise spike not reflected in the noise array) and for the remaining three, we are able to calculate $3\sigma$ lower limits to \Halpha/\Hbeta.
Our final sample then consists of 51 galaxies, for which the Balmer decrement is measured as the ratio between \Halpha and \Hbeta.

\subsection{Stellar mass and star-formation rate measurements}\label{ss.mass_sfr_measurements}

Stellar masses (\Mstar) and star-formation rates (SFR) are obtained again from \ppxf, but using a separate, continuum-focused pipeline. The procedure is described in detail in \citet{looser_jades_2023}; here we report only a brief summary. As input we use the same C3K/MIST SSP library, with a logarithmic grid of 
ages spanning between $10^6$~yr and the age of the Universe at the redshift of the target. The effect of dust is modelled using the Calzetti law \citep{calzetti_dust_2000}. The star-formation history (SFH) is reconstructed using a combination of low regularisation and wild bootstrapping, following the approach of \citet{looser_jades_2023}. Stellar masses are defined as the total mass formed, by integrating the SFH. The SFR is measured as the time-averaged SFH over the last 10~Myr. This value is derived purely from the rest-frame UV stellar continuum (i.e., nebular emission is marginalised over), but \citet{looser_jades_2023} show these measurements trace
the SFR derived combining \Halpha and the Balmer decrement from medium-resolution spectroscopy \citep{curti_jades_2023}, with a scatter of 0.4~dex and an offset of
-0.1~dex. Assuming measurement uncertainties are equal between the \Halpha- and continuum-based SFRs, we obtain a median uncertainty of
0.3~dex on the SFR. Similarly, \citet{looser_jades_2023} derived a median uncertainty on \Mstar of 0.2~dex comparing their measurements to the \Mstar values
from \beagle\ SED models \citetext{\citealp{chevallard_modelling_2016}, Chevallard et~al., in~prep}. Crucially, the \ppxf-derived SFR are not directly coupled with the Balmer decrement (even though we cannot rule out a small correlation due to stellar absorption) therefore we can study the relation between these two galaxy observables with negligible correlated noise, compared to the \Halpha-derived SFR.

\begin{figure}
    \centering
    \includegraphics[width=\columnwidth]{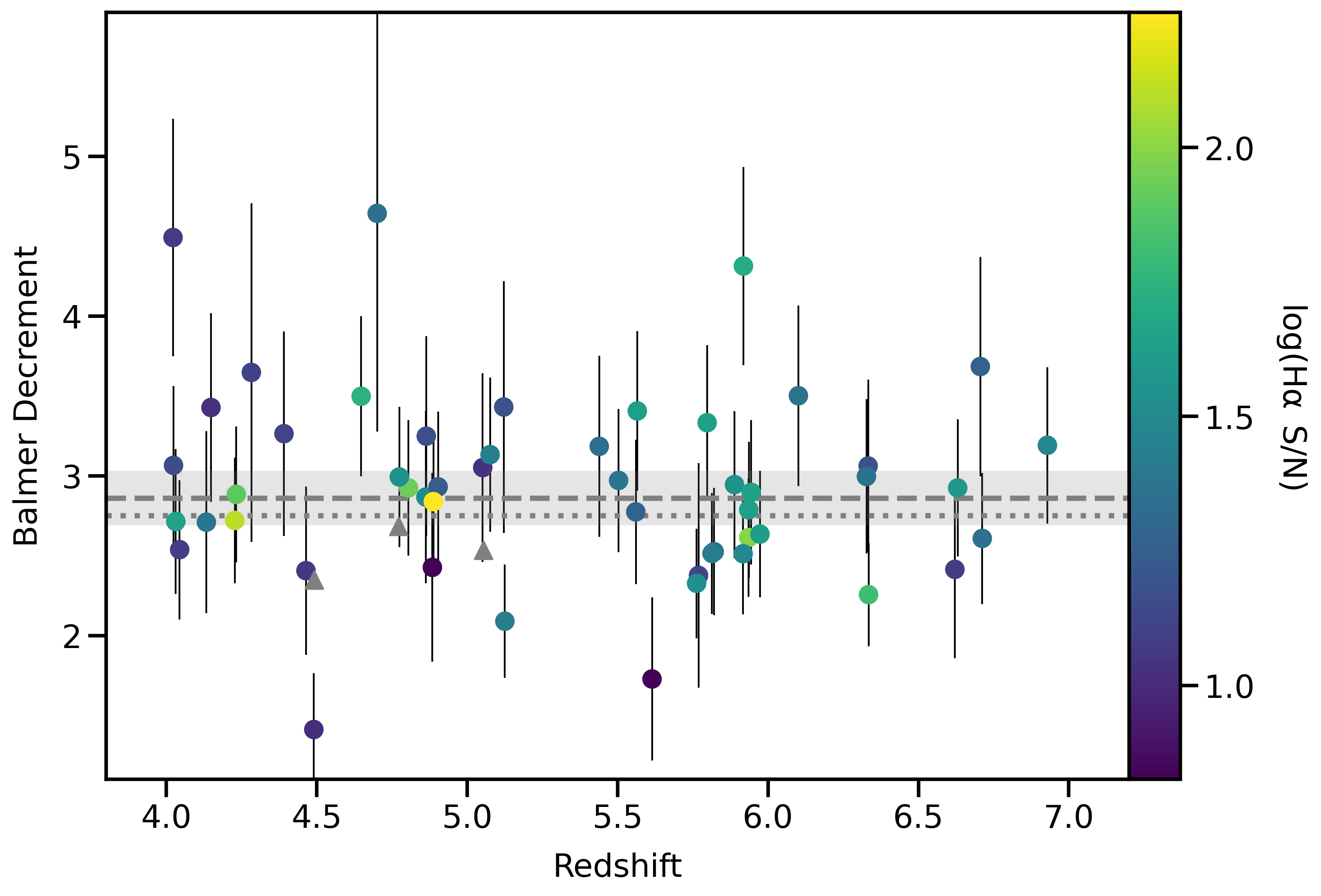}
    \caption{Balmer decrement $\Halpha/\Hbeta$ plotted against redshift for our sample of 51 galaxies. The points are colour coded by their corresponding \Halpha S/N. The errorbars combine the random measurement uncertainties on the line fluxes with a systematic uncertainty of 10 per~cent. Three lower limits are shown as grey triangles. The horizontal dashed grey line corresponds to a Balmer decrement of 2.86, appropriate for local star-forming regions. The grey shaded region represents the 6~per cent range of \Halpha/\Hbeta corresponding to $5,000 < \Telec < 30,000$~K and $10 < \nelec < 500$~\percmq. The horizontal dotted grey line corresponds to a Balmer decrement of 2.75.
    }\label{fig:bd_vs_z.png}
\end{figure}

\section{Results}\label{s.results}

In Fig.~\ref{fig:bd_vs_z.png} we show the Balmer decrement as a function of redshift. Circles represent individual galaxies, colour coded by the S/N of the \Halpha emission line. The errorbars combine the random measurement uncertainties on the line fluxes as well as a systematic uncertainty of 10 per~cent due to flux calibration; grey triangles are $3\sigma$ upper limits. The horizontal dashed line traces the intrinsic (unattenuated) \Halpha/\Hbeta ratio of 2.86, appropriate for Case~B recombination, temperature $\Telec = 10,000$~K and density $\nelec = 100$~\percmq \citep[e.g.,][]{Osterbrock_2006}. Changing these assumptions within the range typical for star-forming regions ($5,000 < \Telec < 30,000$~K and $10 < \nelec < 500$~\percmq) changes the Balmer decrement by 5--6~per cent.

The distribution of measured Balmer decrements is highly scattered, also reaching values below the Case B limit of 2.86. However, within the uncertainties, almost all galaxies below the dashed horizontal line are consistent with a Balmer decrement of 2.86 or higher.
The exception are the two galaxies with the lowest Balmer decrements, which have values clearly deviating from the common assumptions valid in local star-forming regions ($\gtrsim2\sigma$ below the unattenuated value of 2.86). We note, however, that these two galaxies are also the ones with the lowest \Halpha S/N (as highlighted by the colour coding in Fig.~\ref{fig:bd_vs_z.png}), suggesting that the low ratio is due to contamination of \Hbeta and their uncertainties may be underestimated.

\begin{figure*}
    \centering
    \includegraphics[width=\textwidth]{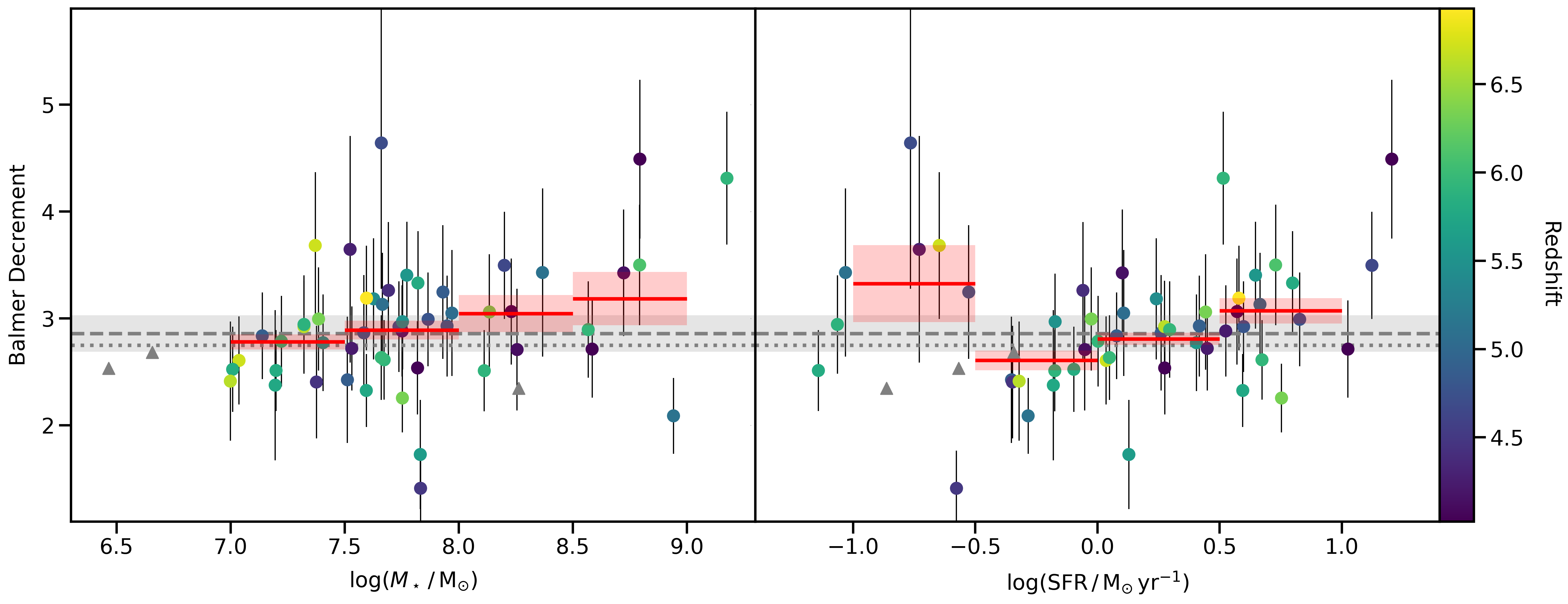}
    \caption{Balmer decrement $\Halpha/\Hbeta$ as a function of \Mstar
    (left panel) and SFR (right panel), both derived from the stellar continuum. Circles represent individual galaxies, colour coded by their spectroscopic redshift. Three lower limits are shown on each panel as grey triangles. The median uncertainties on \Mstar and SFR are
    0.2 and 0.3~dex, respectively.
    The horizontal dashed grey line corresponds to a Balmer decrement of 2.86, appropriate for local star-forming regions. The grey shaded region represents the 6~per cent range of \Halpha/\Hbeta corresponding to $5,000 < \Telec < 30,000$~K and $10 < \nelec < 500$~\percmq. The horizontal dotted grey line corresponds to a Balmer decrement of 2.75. The horizontal red lines and shaded regions represent the
    running mean and error on the mean.
    The Balmer decrement shows statistically significant correlation
    with \Mstar. For SFR, we find an increasing trend from
    $\log (\mathrm{SFR} / \Mstar~\peryr ) > -0.5~$dex. Because
    we measure both \Mstar and SFR from \ppxf (Section~\ref{ss.mass_sfr_measurements}), the \Halpha and \Hbeta emission lines
    do not enter these two measurements, meaning the observed correlations with the Balmer decrement are
    not due to correlated noise.
    }\label{fig:bd_vs_mass_sfr}
\end{figure*}

We do not find a clear correlation with redshift, suggesting that the dust content of galaxies does not evolve significantly between the end of reionisation at $z=7\text{--}6$ and the epoch immediately
before Cosmic Noon ($z=4$), and likely neither the dust properties evolve \citep[consistent with recent findings based on the far-IR/submm properties of galaxies in this redshift range][]{witstok_empirical_2023}.
However, the sample selection criteria prevent us from drawing general conclusions about evolution.

In the left panel of Fig~\ref{fig:bd_vs_mass_sfr} we study the Balmer decrement as a function of \Mstar. We detect a statistically significant Spearman's rank correlation
($r=0.32$, $p<0.05$; traced by the moving mean, horizontal red lines in Fig.~\ref{fig:bd_vs_mass_sfr}), as already found at lower redshifts \citep[e.g.,][]{groves_balmer_2012, shivaei_mosdef_2015, shivaei_mosdef_2020, maheson_unravelling_2023}. We note, however, that the average
Balmer decrement is very low -- particularly so at the low-mass end. For stellar masses $\Mstar < 3\times10^7$~\MSun we find a median decrement of $2.88 \pm 0.08$, very close to the unattenuated value of 2.86 and therefore leaving little room for any dust attenuation. The three lower limits are not constraining (i.e., they are consistent with any amount of dust). This underscores
that -- with the depth of JADES -- we do not miss significant numbers of highly obscured objects due to quality selection cuts.

In the right panel of Fig~\ref{fig:bd_vs_mass_sfr} we study the Balmer decrement as a function of the continuum-inferred SFR, averaged over the last 10~Myr. The symbols and colours are the same as in the left panel. In this case, we do not find a statistically significant Spearman's rank correlation ($r=0.18$, $p>0.05$). However, we do find a 
correlation for galaxies with $\log (\mathrm{SFR} / \Mstar~\peryr ) > -0.5~$dex ($r=0.48$ and $p=0.001$). Note this result is not driven by the high-SFR
outliers: removing also galaxies with $\log (\mathrm{SFR} / \Mstar~\peryr ) > 1$~dex we obtain $r=0.44$ and $p=0.004$.

At this stage, we cannot say whether the Balmer decrement correlates more tightly with \Mstar or with SFR. A comparison of the \ppxf-derived stellar masses to their \textsc{beagle}-inferred equivalents suggests a mean uncertainty of 0.2~dex (accounting for both random noise and systematics due to different SFH assumptions, e.g., \citealt{carnall_how_2019, leja_how_2019, sandles_bayesian_2022}). For the SFR, a comparison between \ppxf and the \Halpha derived values suggests a scatter of 0.4~dex. If we assume that the SFR from \ppxf traces accurately the SFR on 10~Myr timescales, then the measurement uncertainties on SFR are larger than on \Mstar. This would make the \textit{intrinsic} correlation between Balmer decrement
and SFR stronger (i.e., higher correlation coefficient) and more significant than the correlation between Balmer decrement and \Mstar. More data and a better characterisation of the systematic uncertainties could help address which of these two correlations is strongest.

\section{Discussion}\label{s.discussion}

Empowered by the combination of the exquisite sensitivity of \JWST/NIRSpec  with the unprecedented depth of JADES, we can probe for the first time the low-mass end of the galaxy distribution beyond Cosmic Noon, with tens of galaxies in the range $10^7\text{--}10^8$~\MSun.
In this mass regime, we find no evidence of high-attenuation galaxies, consistent with the overall trend of increasing Balmer decrement with increasing \Mstar, already reported by numerous authors for the local Universe \citep{groves_balmer_2012}, Cosmic noon \citep{dominguez_dust_2013, reddy_mosdef_2015, nelson_spatially_2016, lorenz_updated_2023, matharu_first_2023} and redshifts $3\text{--}6$ \citep{shapley_jwstnirspec_2023}.
The lack of objects with significant nebular attenuation is also confirmed by the lack of constraining lower limits.
Part of the reason we find no galaxies with high Balmer decrement could be due to 
sample bias, because our primary targets were selected to be UV-bright 
objects with strong \Lyalpha drops \citet{bunker_jades_2023}.
However, secondary targets are not selected to be as UV bright, therefore selection criteria alone may be insufficient to explain the observed dearth of galaxies with high Balmer decrement. Indeed, our results are consistent with the shallower measurements from the CEERS survey. In the sample of \citet{shapley_jwstnirspec_2023}, there are six galaxies with $5<z<6$ and $\Mstar < 10^8$~\MSun; of these, two imply Balmer decrements larger than 3. The only measurement with a Balmer decrement uncertainty smaller than 1 is fully consistent with our median Balmer decrement in the same mass range.

These findings suggest that, similar to the low-redshift Universe, even in the young
Universe before cosmic noon, more massive galaxies produced more dust and/or could retain more dust. As discussed in \cite{maheson_unravelling_2023}, this might be partly due to the mass-metallicity relation (and the fact that dust content scales with metallicity). Retention of dust due to the gravitational potential seems to also play a role in the mass dependence. However, in addition to these two secondary correlations, dust attenuation is also found to have an intrinsic dependence on stellar mass. This is likely a consequence of the direct correlation between stellar mass and gas mass \citep{lin_almaquest_2019, baker_molecular_2023}. At the high redshifts of our sample, disentangling the dependencies on these individual parameters requires more statistics than currently available (the multi-parameter dependencies found in local galaxies by \citealt{maheson_unravelling_2023} required a sample of thousands of galaxies).

Interestingly, we find evidence of extremely low Balmer decrements at the low-mass end ($2.88\pm0.08$ for $\Mstar < 3\times 10^7$~\MSun). With the physical conditions of local star-forming regions (i.e., $\Telec=10,000$~K and $\nelec=100$~\percmq), the intrinsic Balmer decrement is 2.86. This means our measurement leaves little or no room for \textit{any} ISM dust towards the \HII regions \citep[although dust might be present inside \HII regions, while absorbing UV photons;][]{charlot_star_2002}.
For example, assuming the Milky Way extinction law with $R_V=3.1$ \citep[][hereafter, \citetalias{cardelli_relationship_1989}]{cardelli_relationship_1989} and an intrinsic \Halpha/\Hbeta ratio of 2.86, the observed Balmer decrement yields $A_V=0.02$~mag, meaning very little dust is present. This is a finding that was also inferred by the blue slopes of the spectra of galaxies at such high redshifts \citep{fiore_dusty-wind-clear_2023} and is in agreement with current theoretical models (which predict a drop of the dust-to-stellar mass ratio by a factor of 2 between $z=0$ and $z=4\text{--}7$; \citealp{popping+2017}).

However, other studies have found evidence for some dust attenuation at such high redshifts in low-mass galaxies,
e.g. through the detection of the UV carbonaceous dust absorption dip \citep{markov_dust_2023, witstok_carbonaceous_2023} and through SED modelling \citep[e.g.,][ although these authors
explore a mass at least $2\text{--}3\times$ larger than here]{tacchella_jwst_2023}.

An interesting possibility is that the low Balmer decrement might be associated with `non-standard` physical conditions in the ISM.
In fact, our findings would be consistent with the hypothesis that at
$z=4\text{--}7$, star-forming gas in low-mass galaxies had higher \Telec \citep{schaerer_first_2022, curti_chemical_2023} and/or higher \nelec
\citep{reddy_jwstnirspec_2023}. Qualitatively, this is consistent with lower metallicity stars being the primary source of photoionisation, and with the prediction of higher gas densities from numerical simulations \citep[e.g.][]{ceverino_firstlight_2018, lovell_flares_2022}. For example, assuming $\Telec=20,000$~K and $\nelec=300$~\percmq, the intrinsic \Halpha/\Hbeta ratio drops to 2.75. However, comparing this value to the observed decrement of 2.88, the \citetalias{cardelli_relationship_1989} law gives $A_V=0.14$~mag -- still indicating the presence of only a modest amount of dust.


In any case -- at the low-mass end -- there seems to be very little room for interstellar dust, even taking into account the possible effect of the extreme gas conditions which  might lower the intrinsic Balmer decrement. Part of the difference compared to higher-mass galaxies is the lower metallicity. Several authors \citep{curti_jades_2023, nakajima_jwst_2023} report the mass-metallicity relation is already in place at $z>4$ -- albeit with a large intrinsic scatter. Lower ISM metallicity implies lower dust content in the ISM \citep[especially taking into account that the dust-to-gas ratio drops more than linearly at low metallicities;][]{de_cia_metals_2018}.
In addition, low-mass galaxies could also be more efficient at removing dust.
Numerical simulations \citep{ceverino_firstlight_2018, ma_simulating_2018, lovell_flares_2022, dome_mini-quenching_2023} predict that galaxies with stellar masses $\Mstar \lesssim 0.5\text{--}5 \times 10^8$~\MSun have `burstier' SFHs; their SFR varies rapidly on timescales of 10~Myr, modulated by rapid gas accretion and efficient star formation. This phase is accompanied by gas depletion and strong, efficient feedback which removes any leftover gas \citep[see also discussion in ][]{fiore_dusty-wind-clear_2023}. Burstier SFHs at low mass and high
redshifts seem confirmed by a comparison between the SFRs averaged over 0--10 and 10--90~Myr \citetext{\citealp{looser_jades_2023}, but see \citealp{rezaee_no_2022} for a different view}; our work presents a consistent picture from the point of view of the ISM attenuation. Finally, in terms of dust production timescales, the low dust content may reflect also the fact that low-mass galaxies have younger ages \citep{looser_jades_2023}, hence less time to produce dust through the AGB channel (whereas dust production from SN acts on very short time scales).

Regarding the possible correlation of the Balmer decrement with SFR, we note that the lower metallicity and the burstiness of the SFH require a different calibration of the \Halpha-based SFR indicator, as already recognised by several authors \citep[e.g.,][]{hao_dust-corrected_2011, reddy_hduv_2018, shivaei_infrared_2022, tacchella_h_2022, curti_jades_2023, shapley_jwstnirspec_2023}. Metallicity enters the SFR calibrators through the production rate of hydrogen-ionising photons. Using a solar-metallicity calibrator in a low-metallicity galaxy leads to an overestimate of the SFR by a factor of 2--3.
The burstiness of the SFH implies that the often adopted standard linear relation between \Halpha luminosity and SFR, which assumes constant SFR, cannot be really applied. Additionally, the SFR inferred from \Halpha cannot be used for exploring the relation with the Balmer decrement because the \Halpha flux would enter in both quantities and introduce spurious correlations.

Therefore, we have leveraged on the continuum-based SFR from \ppxf to quantify the correlation between Balmer decrement and SFR (right panel Fig.~\ref{fig:bd_vs_mass_sfr}) and compared it to the well-known correlation with \Mstar (left panel Fig.~\ref{fig:bd_vs_mass_sfr}). The correlation with \Mstar is statistically significant, while for SFR we find a correlation only for SFR$>0.3$~\MSun~\peryr.
We are unable to determine if, for these distant galaxies, any of the two is a secondary correlation, as inferred for local galaxies \citep{maheson_unravelling_2023}. In fact, because of the star-forming main sequence -- i.e., the empirical correlation between \Mstar and SFR -- it is unclear if and to what extent the correlation between Balmer decrement and SFR reflects the combination of the known correlation between Balmer decrement and \Mstar \citep[e.g.,][]{groves_balmer_2012, maheson_unravelling_2023}, and the star-forming main sequence. A larger sample would enable a comparative analysis
using partial correlation coefficients \citep[e.g.,][]{bait_interdependence_2017, bluck_what_2019} or even a machine-learning 
analysis \citep[e.g.,][]{bluck_quenching_2022, piotrowska_quenching_2022}.

Intriguingly, the relation between Balmer decrement and SFR may be tighter than the
relation with \Mstar (except the set of outliers with high Balmer decrement and very low SFR, 0.1--0.3~\MSun \peryr). A primary correlation with SFR would also be
consistent -- at least in a qualitative sense -- with the picture of bursty SFHs
from numerical simulations. Starbursts in low-mass galaxies are fuelled by short-lived reservoirs of dense gas. The burst phase ends as a consequence of the gas being exhausted and/or expelled, therefore low-SFR galaxies would correspond to gas-poor systems observed in a `lull' of their star formation.

The Balmer decrement intrinsically scaling with SFR would also be consistent with other scaling relations observed in the local Universe. Specifically, the dust content scales with the amount of cold gas, via the dust-to-gas ratio. On the other hand, the SFR correlates with the amount of cold gas via the Kennicutt-Schmidt relation. Therefore, a correlation between Balmer decrement and SFR is not unexpected.
In fact, in local galaxies the Balmer decrement correlates with the surface density of molecular gas \citep{barrera-ballesteros_sdss-iv_2018, piotrowska_towards_2020}.

Other works \citep{maheson_unravelling_2023} have argued that -- after controlling for inclination -- the correlation between SFR and Balmer decrement is less important than the correlation of Balmer decrement with \Mstar and other physical quantities (ISM metallicity and gas velocity dispersion).
However, they are unable to compare the Balmer decrement to the SFR on timescales of 10~Myr -- simply because the classic \Halpha-based estimator uses the Balmer decrement itself in its definition.
Instead, they use the 4000~break as a proxy for the current SFR. However, the 4000~break is sensitive to SFR on longer timescales of 100~Myr, so their different results could also be due to different timescales being probed.
In any case, their much more detailed analysis is impossible with our sample size of only 51 galaxies, but requires samples 10--100 times larger that do not exist yet for the redshift range we are interested in.

A possible caveat is represented by the effect of different morphologies and inclination. We do not make any morphology cut, and assume that our sample probes all morphologies and inclinations equally. Larger samples may help clarify the role of morphology and inclination in our findings.

\section{Summary and Conclusions}\label{s.conclusions}

In this work, we presented a comparative analysis of the Balmer decrement for galaxies with $\Mstar = 10^7\text{--}10^9$~\MSun between cosmic noon and the end of
reionisation ($4 < z < 7$). We used deep (28~hours), low spectral resolution 
observations of \JWST/NIRSpec MSA, obtained as part of JADES. We use medium-resolution observations to apply a correction for \NII contamination of \Halpha. We further explore the dependence on SFR, averaged over the last 10~Myr, as measured from the stellar continuum -- without using information from the emission lines.

We confirm earlier findings of a correlation between Balmer decrement (i.e. dust attenuation) with \Mstar, already in place at such early epochs.

At the low-mass end ($\Mstar < 3\times10^7$~\MSun), there is little evidence of any dust attenuation, in agreement with theoretical expectations \citep{popping+2017}.
This result is consistent with the expectation from these systems being metal poor, young (hence with less time for dust formation) and with bursty SFHs (which may be more effective in depleting and expelling dust). In this mass range, the median Balmer decrement is only $2.88\pm0.08$, extremely close to 2.86 (the Case B value commonly assumed for star-forming regions at lower redshifts). 
This suggests that low-mass galaxies at these redshifts might have higher \Telec and/or \nelec compared to their local counterparts, which would result into a slightly lower intrinsic Balmer decrement.

We find tentative evidence of a correlation between the Balmer decrement and SFR. 
If confirmed, this may trace the underlying correlation between SFR and cold gas mass, via the dust-to-gas ratio.


Future studies based on larger samples may clarify which of these correlations are intrinsic and direct, and which ones are an indirect byproduct of secondary correlations. Larger samples will also be able to uncover correlations with other galaxy parameters, such as metallicity and velocity dispersion \citep{maheson_unravelling_2023}.

\section*{Acknowledgements}

LS, FDE, RM and TJL acknowledge support by the Science and Technology Facilities Council (STFC) and ERC Advanced Grant 695671 "QUENCH". RM also acknowledges funding from a research professorship from the Royal Society.



\bibliographystyle{config/aa}
\bibliography{bib_006} 








\label{lastpage}
\end{document}

%% file: config/macros.tex
\usepackage{caption}
\usepackage{hyperref}
\hypersetup{
    colorlinks=true,
    linkcolor=blue,
    citecolor=blue,
    filecolor=magenta,
    urlcolor=cyan,
    }
\usepackage{amsmath}
\usepackage{subcaption}
\usepackage{slantsc}
\usepackage{textgreek}
\usepackage{xargs}
\usepackage{xcolor}
\usepackage{xspace}

\newcommand{\oii}{\relax \ifmmode {\mbox O\,{\scshape ii}}\else O\,{\scshape ii}\fi}
\newcommand{\oiii}{\relax \ifmmode {\mbox O\,{\scshape iii}}\else O\,{\scshape iii}\fi}
\newcommand{\Lyalpha}{\text{Ly\textalpha}\xspace}
\newcommand{\Halpha}{\text{H\textalpha}\xspace}
\newcommand{\Hbeta}{\text{H\textbeta}\xspace}

\newcommand{\NII}{\text{[N\,{\sc ii}]}\xspace}
\newcommandx{\NIIL}[1][1=6584]{\text{[N\,{\sc ii}]\textlambda {#1}}\xspace}

\newcommand{\OIII}{\text{[O\,{\sc iii}]}\xspace}
\newcommandx{\OIIIL}[1][1=5007]{\text{[O\,\textsc{iii}]\textlambda{#1}}\xspace}

\newcommandx{\altOIIIL}[1][1=5007]{\text{[O\,{\footnotesize III}]\textlambda{#1}}\xspace}

\newcommandx{\CIVL}[1][1=1550]{\text{C\,{\sc iv}\textlambda{#1}~\AA}\xspace}

\newcommandx{\OIIIsL}[1][1=1665]{\text{O\,{\sc iii}]\textlambda{#1}~\AA}\xspace}

\newcommandx{\CIIIL}[1][1=1908]{\text{C\,{\sc iii}]\textlambda{#1}~\AA}\xspace}

\newcommandx{\HeIIL}[1][1=1640]{\text{He\,{\sc ii}\textlambda{#1}~\AA}\xspace}
\newcommand{\HII}{\text{H\,{\sc ii}}\xspace}

\definecolor{ForestGreen}{HTML}{2e8b21}
\definecolor{FireBrick}{HTML}{b22222}

\let\oldtextsigma\textsigma
\renewcommand{\textsigma}{\oldtextsigma\xspace}
\let\oldtextalpha\textalpha
\renewcommand{\textalpha}{\oldtextalpha\xspace}
\let\oldAA\AA
\renewcommand{\AA}{\text{\oldAA}\xspace}

\newcommand{\Mstar}{\ensuremath{M_\star}\xspace}
\newcommand{\MSun}{\ensuremath{{\rm M}_\odot}\xspace}

\newcommand{\peryr}{\ensuremath{{\rm yr^{-1}}}\xspace}
\newcommand{\percmq}{\ensuremath{{\rm cm^{-3}}}\xspace}

\newcommand{\Telec}{\ensuremath{T_{\rm e}}\xspace}
\newcommand{\nelec}{\ensuremath{n_{\rm e}}\xspace}

\defcitealias{cardelli_relationship_1989}{CCM89}
\defcitealias{shapley_jwstnirspec_2023}{S23}

\newcommand{\JWST}{\textit{JWST}\xspace}